\documentclass[sigconf]{acmart}
\usepackage{subfigure}
\usepackage{multirow}
\usepackage{enumitem} 
\usepackage{graphics}
\usepackage{pifont}
\newtheorem{theorem}{Theorem}
\AtBeginDocument{%
  \providecommand\BibTeX{{%
    \normalfont B\kern-0.5em{\scshape i\kern-0.25em b}\kern-0.8em\TeX}}}

\setcopyright{acmcopyright}
\copyrightyear{2018}
\acmYear{2018}
\acmDOI{10.1145/1122445.1122456}

\acmConference[Woodstock '18]{Woodstock '18: ACM Symposium on Neural
  Gaze Detection}{June 03--05, 2018}{Woodstock, NY}
\acmBooktitle{Woodstock '18: ACM Symposium on Neural Gaze Detection,
  June 03--05, 2018, Woodstock, NY}
\acmPrice{15.00}
\acmISBN{978-1-4503-XXXX-X/18/06}

\begin{document}

\settopmatter{printacmref=false}
\title[The BROOK Dataset]{BROOK Dataset: A Playground for Exploiting Data-Driven Techniques in Human-Vehicle Interactive Designs}


\author{Wangkai Jin, Yicun Duan, Junyu Liu, Shuchang Huang, Zeyu Xiong, Xiangjun Peng}
\affiliation{
    \institution{User-Centric Computing Group,University of Nottingham Ningbo China}
    \country{https://unnc-ucc.github.io/index.html}
}


\begin{abstract}
  Emerging Autonomous Vehicles (AV) breed great potentials to exploit data-driven techniques for adaptive and personalized Human-Vehicle Interactions. However, the lack of high-quality and rich data supports limits the opportunities to explore the design space of data-driven techniques, and validate the effectiveness of concrete mechanisms. Our goal is to initialize the efforts to deliver the building block for exploring data-driven Human-Vehicle Interaction designs. To this end, we present BROOK dataset, a multi-modal dataset with facial video records. We first brief our rationales to build BROOK dataset. Then, we elaborate how to build the current version of BROOK dataset via a year-long study, and give an overview of the dataset. Next, we present three example studies using BROOK to justify the applicability of BROOK dataset. We also identify key learning lessons from building BROOK dataset, and discuss about how BROOK dataset can foster an extensive amount of follow-up studies.

\end{abstract}

\maketitle

\section{Introduction}
Recent advances in data-driven techniques (e.g. Neural Networks) breed an extensive amount of opportunities, to enable adaptive and personalized Human-Vehicle Interaction (HVI). More interestingly, the emerging trends of Autonomous Vehicles relax the conventional burdens of driving, and in-vehicle drivers/passengers are capable to obtain better user experiences through more complex HVI. Incorporating data-driven approaches can greatly improve user experiences during driving processes. For instance, unobtrusive monitors of multi-modal statuses (i.e. by taking alternative data sources as input), rather than directly equipping biosensors, are more user-friendly as data sources for personalized interaction between drivers and vehicles. Therefore, combining data-driven techniques with HVI becomes promising in the near future, and relevant datasets become essential and highly demanded. 

Though there are already several datasets for HVI, there are four major limitations of existing datasets. First, the design purposes of existing datasets focus on a specific type of study purposes (e.g. Stress Detection \cite{MITDB}, Driving Workload \cite{HciLabDB} and etc.), which naturally limits their potentials due to the narrowly-selected types of data streams. Second, existing datasets only collect information when drivers are in manual mode, which is not suitable for the emerging trends of Autonomous Vehicles. Third, existing datasets don't have sufficient coverage of the contextual data, especially the statistics of Vehicle Statuses; and fourth, existing datasets don't account for advanced bio-sensors (e.g. Eye-Tracking devices), which restricts the volumes of physiological data (Section~\ref{sec:background}).  

Therefore, a comprehensive dataset can unleash the opportunities to leverage data-driven designs for HVI and validate their effectiveness in practice. Data-driven designs usually demand large-scale datasets as the building block, for training, testing and validating their performance and efficiency. Our goal in this work is to initialize constructing such a building block to stimulate diverse types of explorations and allow reasonable comparisons across different designs. We believe that achieving such a goal is two-folded. First, we aim to build such a dataset as comprehensively as possible; and second, we aim to justify the applicability of such a dataset via concrete demonstrations of example studies.

To this end, we present BROOK\footnote{The name "BROOK" refers to multiple water flows, which stands for the key idea of our dataset} dataset, a multi-modal and facial video dataset for data-driven HVI. The aim of the BROOK dataset is to provide a \textbf{publicly available}\footnote{BROOK dataset is approved to be shared by all participants, would be released on a public website, and would be accessed upon requests.} and \textbf{large-scale} dataset within the domain of HVI. The outcome of building the BROOK dataset is the most comprehensive dataset in HVI to date, which contains much more kinds of data streams than previous datasets. The key idea, behind BROOK, is to \textbf{map as many kinds of data flows as possible in the same timeline}, when participants are under both manual and fully-automated driving scenarios. Such guiding philosophy allows BROOK to retrieve statistics from 34 drivers in 11 dimensions (i.e. 4 physiological and 7 contextual). We provide sufficient details regarding how BROOK is organized, and how we build the BROOK dataset, via a year-long study. We also provide detailed descriptions and release the sources of all relevant software, when we are building the BROOK dataset\footnote{Software supports for the BROOK dataset would be released simultaneously with the release of the dataset.} (Section~\ref{sec:BROOK}).

To justify the applicability of the BROOK dataset, we design and perform three example studies to demonstrate the potentials of the BROOK dataset. First, we provide unobtrusive techniques to collect multi-modal statistics from drivers, without equipping biosensors. We achieve this through Convolution Neural Networks over the BROOK dataset, by taking facial expressions as the only input. Second, we quantitatively demystify complicated interactions between driving behaviors and driving styles and provide an adaptive approach to handle such complexity. We achieve this through Self-Clustering algorithms over the BROOK dataset, by dynamically identifying the number of clusters and merging data points accordingly. Third, we identify the tradeoffs between different privacy-preserving techniques in the context of Internet-of-Vehicles. By exploiting four state-of-the-art Differential Privacy mechanisms, we retrieve key observations in terms of accuracy-privacy tradeoffs among different mechanisms. We believe the BROOK dataset can be applied to more contexts beyond the above three examples, by taking different types of data streams as primary sources (Section~\ref{sec:example}).

We highlight the most significant results/findings from our three example studies. For our first example. we demonstrate great potential for this unobtrusive alternative, by achieving 82.96\%, 57.90\%, and 58.75\% accuracy in terms of Skin Conductance, Heart Rates, and Vehicle Speed predictions (i.e. by taking facial videos as the only input, rather than equipping the sensors). For our second example, we justify that Self-Clustering algorithms can quantitatively classify diverse combinations of different driving styles in an adaptive manner. And for our third example, we identify the needs of co-designing the applications and privacy-preserving techniques for data utility and privacy, in the context of Internet-of-Vehicles. The above results suggest the potentials of both the BROOK dataset and example studies are significant, and we conjecture that future characterizations and explorations of BROOK are promising. 

We position the BROOK dataset as a starting point for data supports of adaptive and personalized Human-Vehicle Interactions, rather than a definitive complete answer. This is because our study has several limitations, which are incapable to be resolved by a single research group. We envision that a community-wide collaboration is essential for building a complete dataset. We discuss the potential optimizations of our dataset and summarize key learning lessons for future attempts, to provide a comprehensive dataset for HVI (Section~\ref{sec:futureresearch}). 

We make the following three key contributions in this paper:

\begin{itemize}
    \item We present the BROOK dataset, a multi-modal and facial video dataset which consists of 34 real-world drivers and their statistics in 11 dimensions. The coverage of BROOK includes driving statuses, multi-modal statistics and video recordings during both manual and autonomous driving procedures. 
    \item We present three example studies, to demonstrate the potentials of the BROOK dataset. Our empirical results confirm the quality of the BROOK dataset, and suggest novel insights for using data-driven techniques in Human-Vehicle Interaction. 
    \item We summarize important lessons when building and using BROOK dataset, and discuss potential directions for more usages of the BROOK dataset. We hope that these lessons can help with future efforts to build or leverage BROOK-like datasets for Human-Vehicle Interaction.
\end{itemize}

\noindent
The rest of this paper is organized as follows. Section~\ref{sec:background} presents the background and motivation to build BROOK dataset. Section~\ref{sec:BROOK} gives an overview and present details on how we collect BROOK dataset. Section~\ref{sec:example} presents three example studies, including: (1) an unobtrusive technique to obtain multi-modal and driving statistics via only facial expressions; (2) our design, implementation and empirical evaluations to demystify the complicated interactions between driving behaviors and styles; (3) our design, deployment and characterization of different methods of Differential Privacy for Internet-of-Vehicles. Section~\ref{sec:futureresearch} covers limitations and lessons we learned while building the BROOK dataset, and discusses potential research directions using BROOK-like datasets.

\section{Background \& Motivation}
\label{sec:background}

In this section, we provide background and motivation to build BROOK dataset. We first identify the growing importance of data-driven techniques in Human-Vehicle Interaction (Section~\ref{subsec:data-driven}). Then, we briefly elaborate existing datasets and their usage scenarios for Human-Vehicle Interaction, and identify their design purposes accordingly (Section~\ref{subsec:existing-datasets}). Finally, we comprehensively compare the detailed compositions of existing datasets and our proposed BROOK dataset, to justify our novelty (Section~\ref{subsec:exist_dataset_limitation}).

\subsection{Data-Driven Techniques for Human-Vehicle Interaction}
\label{subsec:data-driven}

With the emerging trend of Autonomous Vehicles in the industry, research interests have grown significantly in Human-Vehicle Interaction (HVI) field, trying to tackle various problems for the coming era. A recent summary categorizes existing studies into two general types, according to manual and autonomous driving contexts respectively \cite{HVIoutlook}. The underlying essence for the HVI study is to understand both drivers' states and the temporal events outside the vehicle, by leveraging multiple data-driven techniques. Current data-driven techniques are designed for various purposes, such as (1) tracking driver attention \cite{MonitorAttention1,MonitorAttention2}; (2) analyzing driving styles \cite{AnalyzeDS1,AnalyzeDS2,AnalyzeDS3}; (3) analyzing effects of situations outside the vehicle \cite{OutsideSituation1,OutsideSituation2,OutsideSituation3};(4) predicting drivers' drowsiness, attention, and emotion through facial expressions or eye gazes \cite{DMD,DRVE,dada,3MDAD}; (5) monitoring driver's behaviors regarding distractions and temporal performance by utilizing body-based monitoring mechanisms \cite{Body1,Body2}; (6) detecting drivers' emotion arousal (e.g. neutral, sadness and happiness) \cite{emotion1,emotion2,emotion3}. To sum up, current data-driven techniques for HVI study have introduced valuable insights towards adaptive and personalized HVI systems so far, yet it is essential to explore more advanced techniques for the field with more diversified data inputs.

\subsection{Existing Datasets for Human-Vehicle Interaction}
\label{subsec:existing-datasets}
There are several publicly available datasets for advanced Human-Vehicle Interaction Research. Hereby, we summarize five key datasets and discuss their application scenarios. Table~\ref{tab:design-purposes} presents the comparison between existing datasets and BROOK dataset (our proposal). Distracted Driver Dataset \cite{DistractedDriver} aims to provide a solid foundation for studies regarding stress detection and posture classification, during the driving procedure; brain4cars \cite{brain4cars} aims to enable design exploration in terms of Maneuver Estimation; MIT driverDB \cite{MITDB} provides relatively more types of data streams for detailed characterizations of stress detection; HciLab \cite{HciLabDB} aims to measure driving workloads and provides empirical studies for proofs-of-concept; and AffectiveROAD \cite{AffectiveROAD} targets personalized Human-Vehicle Interactions, by providing relevant monitors of affective states. All the above datasets are purpose-dependent and there are certain limitations for exploring more design spaces since their design purposes are deterministic and centralized on particular domains. Different from the above datasets, the goal of BROOK is to provide a comprehensive dataset to exploit data-driven techniques for Adaptive and Personalized Human-Vehicle Interaction.

\begin{table*}
\caption{A comparison between BROOK dataset and existing datasets in terms of design purposes.}
\begin{tabular}{|c|c|}
\hline
\textbf{Dataset} & \textbf{Design Purposes} \\
 &  \\ \hline
  &  \\ 
\textbf{Distracted Driver Dataset} \cite{DistractedDriver} & \begin{tabular}[c]{@{}c@{}}Stress Detection \& Posture classification\end{tabular} \\ 
 &  \\ \hline
  &  \\ 
\textbf{brain4cars} \cite{brain4cars} & Maneuver Estimation \\ 
 &  \\ 
\hline
 &  \\ 
\textbf{MIT driverDB} \cite{MITDB}& Stress Detection \\ 
 &  \\ 
 \hline
  &  \\ 
\textbf{HciLab} \cite{HciLabDB} & \begin{tabular}[c]{@{}c@{}}Driving Workload Measurement\end{tabular} \\ 
 &  \\ 
 \hline
  &  \\ 
\textbf{AffectiveROAD} \cite{AffectiveROAD} & \begin{tabular}[c]{@{}c@{}}Affective State Monitoring\end{tabular} \\ 
 &  \\ 
 \hline
  &  \\ 
\underline{\textbf{BROOK}} (Our Dataset)& \begin{tabular}[c]{@{}c@{}}\underline{Data-driven Techniques in Human-Vehicle Interaction}\end{tabular} \\ 
 &  \\ 
 \hline
\end{tabular}%

    \label{tab:design-purposes}
\end{table*}

\subsection{Detailed Comparisons among Existing Datasets and BROOK Dataset}
\label{subsec:exist_dataset_limitation}
We first brief different types of data streams within the above datasets and the BROOK dataset. Table~\ref{tab:comparison} presents the detailed comparisons between existing datasets and the BROOK dataset. Compared to the BROOK dataset, existing datasets share two major limitations: (1) most datasets are designed for study-specific purposes, which might be restrictive for exploring novel designs; and (2) most datasets focus on the monitoring of drivers' physiological states and analysis of affective states, but the important statistics of driving context are relatively overlooked. As Table~\ref{tab:comparison} shows, existing publicly available datasets fall short of contextual information, which would be critical for the development of adaptive and personalized HVI systems for complicated driving scenarios.

\begin{table*}
\caption{A comparison, in terms of contained data types, between BROOK dataset and existing datasets for Human-Vehicle Interaction. Those types, which can only be accessed from BROOK, are underlined. Note that EDA stands for Electrodermal Activity, ECG stands for Electrocardiogram, EMG stands for Electromyogram, and HR stands for Heart Rate.}
\resizebox{\textwidth}{!}{%
\begin{tabular}{|c|c|c|c|c|}
\hline
\multirow{2}{*}{} & \multirow{2}{*}{\begin{tabular}[c]{@{}c@{}}\textbf{Number of} \\ \textbf{Drivers}\end{tabular}} & \multirow{2}{*}{\begin{tabular}[c]{@{}c@{}}\textbf{Driving}\\ \textbf{Mode}\end{tabular}} & \multicolumn{2}{c|}{\textbf{Data Types}} \\ \cline{4-5} 
 &  &  & \begin{tabular}[c]{@{}c@{}}\textbf{Driver} \\ \textbf{Physiological Data}\end{tabular} & \textbf{Contextual Data} \\ \hline

\begin{tabular}[c]{@{}c@{}}\textbf{Distracted} \\ \textbf{Driver} \\ \textbf{Dataset} \cite{DistractedDriver}\end{tabular} & 31 & Manual & 10 classes of driver postures & / \\ \hline
\textbf{brain4cars} \cite{brain4cars} & 10 & Manual & Facial Video & \begin{tabular}[c]{@{}c@{}}GPS,\\ Vehicle Speed, \\ Road Video\end{tabular} \\ \hline
\textbf{MIT driverDB} \cite{MITDB} & 24 & Manual & \begin{tabular}[c]{@{}c@{}}EDA,\\ ECG,\\ EMG,\\ HR,\\ Facial Video\end{tabular} & Road Video \\ \hline
\textbf{HciLab} \cite{HciLabDB} & 10 & Manual & \begin{tabular}[c]{@{}c@{}}EDA, \\ ECG, \\ Body Temperature,\\ Facial Video\end{tabular} & \begin{tabular}[c]{@{}c@{}}In-vehicle Brightness, \\ Vehicle Acceleration,\\ Road Video\end{tabular} \\ \hline
\textbf{AffectiveROAD} \cite{AffectiveROAD} & 10 & Manual & \begin{tabular}[c]{@{}c@{}}EDA, \\ Body Temperature,\\ HR, \\ Hand Movement, \\ Facial Video\end{tabular} & \begin{tabular}[c]{@{}c@{}}Luminance, \\ Temperature,\\ GPS,\\ Pressure,\\ Humidity, \\ Sound,\\ Road Video\end{tabular} \\ \hline
\begin{tabular}[c]{@{}c@{}}\textbf{BROOK}\\(Our Dataset)\end{tabular} & 34 & \begin{tabular}[c]{@{}c@{}}Manual,\\\underline{\textbf{Automated}}\end{tabular} & \begin{tabular}[c]{@{}c@{}}HR,\\ EDA, \\ \textbf{\underline{Eye-Tracking}},\\ Facial Video\end{tabular} & \begin{tabular}[c]{@{}c@{}}Vehicle Speed,\\ Vehicle Acceleration,\\ Vehicle Coordinate,\\ Distance of Vehicle Ahead, \\ \textbf{\underline{Steering Wheel Coordinates}},\\ \textbf{\underline{Throttle Status}}, \\\textbf{\underline{Brake Status}}, \\Road Video\end{tabular} \\ \hline
\end{tabular}%
}
\label{tab:comparison}
\end{table*}

We describe the compositions of existing datasets in detail. Most of them are designed to study drivers' affective states during driving procedures. The MIT driverDB  dataset \cite{MITDB} collects drivers' physiological data (Electrodermal Activity (EDA), Electrocardiogram (ECG), Electromyogram (EDM) and etc.), together with in-vehicle and in-field videos from 20 miles of driving between the city and the highway (i.e. in the Greater Boston area). HciLab dataset \cite{HciLabDB} has similar driver physiological data (EDA,ECG, Body Temperature and Facial Video) and different contextual data (In-vehicle Brightness, Vehicle Acceleration, and Road Video), which are collected from 10 drives in highway and freeway in Germany. AffectiveROAD dataset \cite{AffectiveROAD} obtains its data from 13 drives performed by 10 drivers, which covers EDA, Body Temperature, Heart Rate, Hand Movement, and Facial Video for driver statistics, with multiple situational data, such as Luminance, Temperature, Pressure, Humidity, etc. Both HciLab dataset and AffectiveROAD dataset include a self-scored, self-reported stress level from the driver after the experiments. A similar dataset to AffectiveROAD, brain4cars \cite{brain4cars}, contains drivers' facial video,  road video, vehicle GPS information and a speed logger about 10 drivers with different kinds of driving maneuvers across two states, who drive their private cars for 1180 miles between natural freeway and cities. Distracted Driver dataset \cite{DistractedDriver} mainly contains 10 different types of  driver postures, such as safe driving, talking to the phone, adjusting radio, etc., which are collected from 31 participants. 

Compared with existing datasets, the BROOK dataset has three major differences: (1) the above datasets are mainly designed for driver stress level monitoring, while BROOK can support more diverse applications such as driver-style classifications; (2) the above datasets only contain manual mode driving data, while BROOK has both manual mode and auto mode driving statistics; and (3)  BROOK collects more types of vehicle data (e.g. steering wheel positions, brake and throttle status) from more drivers (i.e. 34 drivers in total), which not only provides more contextual information in the driving process but also guarantees the diversity of driving statistics.

\section{BROOK Dataset}
\label{sec:BROOK}
In this section, we introduce BROOK dataset in detail. We first give an overview of BROOK dataset, regarding its high-level characteristics and compositions. Then we introduce the hardware and software for collecting BROOK dataset. Next, we highlight novel features under our consideration, when we are building BROOK dataset. Finally, we present the procedure of data collection for BROOK dataset.

\subsection{An Overview of BROOK Dataset}

\begin{figure*}
    \centering
    \includegraphics[width = \textwidth]{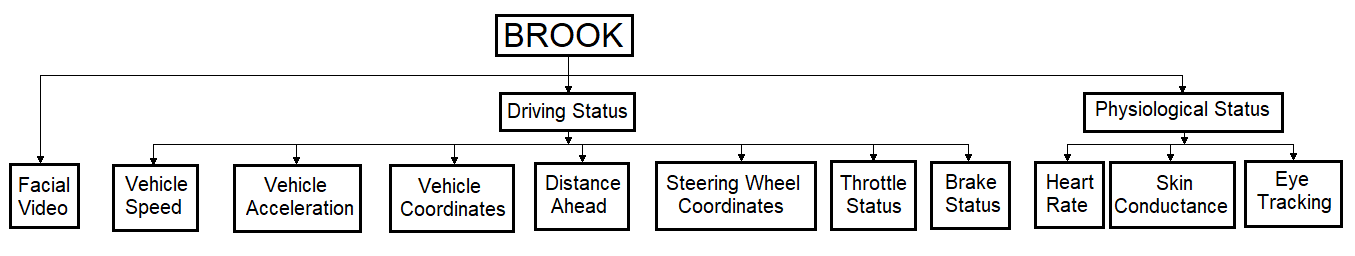}
    \caption{The Composition of the Current BROOK dataset.}~\label{structure}
\end{figure*}

The current version of BROOK dataset has facial videos and many multi-modal/driving status data of 34 drivers. Statistics of each participant are collected within a 20-minute in-lab studies, which is separated into manual mode and automated mode separately. Figure~\ref{structure} shows 11 dimensions of data for the current BROOK dataset, which includes: \textbf{Facial Video, Vehicle Speed, Vehicle Acceleration, Vehicle Coordinate, Distance of Vehicle Ahead, Steering Wheel Coordinates, Throttle Status, Brake Status, Heart Rate, Skin Conductance, and Eye Tracking}.


We follow the guiding principle while building the BROOK dataset, so that we map all kinds of data streams to the same time flow. To summarize our collection, we use the relationships between data types and data sizes for demonstration. Table~\ref{tab:summary} shows the main characteristics of BROOK. Note that Driving Status includes Vehicle Speed, Vehicle Acceleration, Vehicle Coordinate, Distance of Vehicle Ahead, Steering Wheel Coordinates, Throttle Status and Brake Status.

\begin{table*}
\caption{A Summary of the Current BROOK dataset.}~\label{BROOKOverview}
\begin{tabular}{|c|c|c|c|c|c|}
\hline
Data Types & Facial Video & Heart Rates & Skin Conductance & Eye-tracking & Driving Status \\ \hline
Size       & 273GB        & 13.1MB      & 48.3MB           & 27.2GB       & 60.9MB         \\ \hline
\end{tabular}

\label{tab:summary}
\end{table*}

\subsection{Hardware and Software Supports}

\begin{figure}
    \centering
    \includegraphics[width=.8\linewidth]{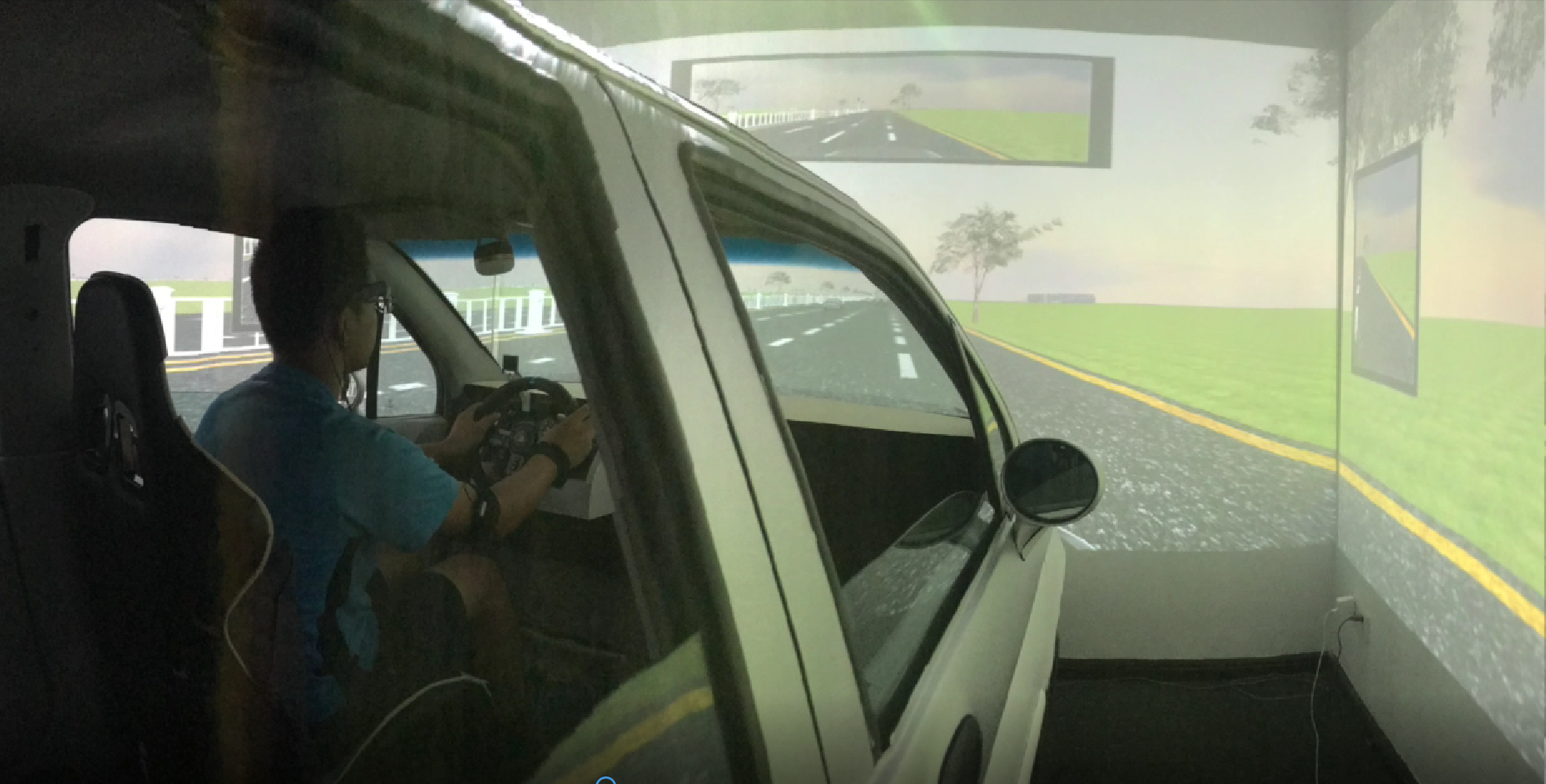}
    \caption{The simulator environment for data collection}
    \label{fig:simulator}
\end{figure}

To build the BROOK dataset, we combine a diverse set of hardware and software and make intensive implementations to facilitate our needs. The simulator environment is showed in Figure~\ref{fig:simulator}. We use an empty car skeleton with replaced steering wheel, pedals and cameras for data collection and integrate them into our simulator software. The camera is positioned on top of the dashboard, in front of the driver's face. The simulated scenes are projected onto three surrounding walls of the car, with the front wall showing the driving scenes ahead of the car and the left/right walls showing the side views and backward views. For data collection, we use different sensors to collect relevant data streams: (1) for Facial Video, we use an ultra-clear motion tracking camera; (2) for Hear Rates,  we use a heart rate sensor; (3) for Skin Conductance, we use a leather electrical sensor; and (4) we use a Tobii eye tracker for Eye-tracking. Detailed sensors types and sample rates are showed in Table~\ref{tab:device}. As for detailed driving status, pressure sensors for throttle and brake have been integrated, and steering wheels have been monitored to obtain spatial changes from drivers during the whole period. 

\begin{table}[!h]
\caption{Details about different sensors used when collecting BROOK dataset.}
\begin{tabular}{|c|c|c|}
\hline
\textbf{Data Types}                     & \textbf{Device Type}                   & \textbf{Frequency} \\ \hline
Facial Video                            & Goldensky Webcam                       & 30Hz        \\ \hline
Heart Rate                              & POLAR H10 Heart Rate Monitor           & 21Hz        \\ \hline
Skin Conductance                        & NeXus Skin Conductance Sensor          & 40Hz   \\ \hline
\multicolumn{1}{|c|}{Eye-Tracking Data} & \multicolumn{1}{c|}{Tobii eye tracker} & 60Hz        \\ \hline
\end{tabular}

\label{tab:device}
\end{table}

As for software, we use OpenDS \cite{OpenDS,OpenDS-Official}, an open-source cognitive driving simulator, for driving scenario generations. OpenDS is widely-used for in-lab simulations for both manual and automated driving. Also, we utilise ErgoLAB \cite{ErgoLab} to coordinate sensor data output for the synchronization and the storage. Finally, we build an in-framework tool to stream out speed and coordinates as driving status.

\subsection{Key Features Only in BROOK}

BROOK has some outstanding features, compared with existing datasets. Hereby, we summarize all three unique features in BROOK in the following paragraphs, which are Automated Mode, Region-based Customization and Eye-Tracking Statistics.

\noindent
\textbf{\underline{1. Automated Mode.}} As shown in Table~\ref{tab:comparison}, BROOK has collected all types of data streams from both manual and automated driving processes. For automated driving, the reported statistics are collected during fully-automated driving processes for all participants. This is because,  as the emerging trends of Autonomous Vehicles (AV), it's important to supply relevant data streams for different studies. For example, \cite{Auto-trust} suggests the need for extensive characterizations of drivers' perceived trust in AV.\\

\noindent
\textbf{\underline{2. Region-based Customization.}} To provide comprehensive coverage of all possible cases, we customize our scenarios \cite{OpenDS-Scenarios} and scenes \cite{OpenDS-Scenes} based on the common patterns of roads in China, as shown in Table \ref{RoadPatterns}. These patterns are derived from our collaborations with local transportation authorities. During the procedure of data collection, all participants have experienced 4 types of driving scenarios. Note that all participants during this study have held issued driving licenses from transportation authorities, with sufficient driving experiences. \\

\begin{table}[h]
    \centering
    \caption{The configurations of experimental driving scenarios and scenes.}~\label{RoadPatterns}
    \begin{tabular}{c c c}
      { \textbf{}} & { \textbf{The Number of of Lanes}}
      & { \textbf{The Number of Cars}} \\
      \toprule   
       1st & 4 & 32\\
      \midrule
       2nd & 4 & 60\\
      \midrule
       3rd & 8 & 60\\
      \midrule
       4th & 8 & 92\\
      \bottomrule
    \end{tabular}
    
\end{table}

\noindent
\textbf{\underline{3. Eye-Tracking Statistics.}} We have collected drivers' Eye-Tracking statistics via a Tobii Eye-Tracking device. Eye-Tracking analysis leverages the fixations and pupil-size variation of each driver, to study diverse types of drivers' behaviors. It's also a widely-recognized assumption that fixations and pupil-size variations are important indicators for driver attention and emotional states. In addition, the Eye-Tracking statistics can be combined with road videos to locate region-of-interests/object-of-interests in the scenes, which can not only profile drivers' body movements in a new perspective but also provide important data to improve the situation awareness of Human-Vehicle Interaction systems.

\subsection{Data Collection Procedure}

In this section, we illustrate the whole procedure while collecting the BROOK database. We first prepare necessary resources and tests to ensure the bulk of the study runs smoothly. Then we perform the bulk of the study to collect necessary data streams, in various settings and modes separately.

For preparations, participants are first briefed about the study and are screened by a questionnaire to ensure that, they are at low risk of motion sickness while experiencing the driving simulation. After the introduction, participants are given five minutes to sit in the simulator by themselves and adapt to the environment. The study only begins when we ensure that participants understand the brief and can operate the simulator without any questions or difficulties.

The bulk of this study is as follows. The experiment includes three driving conditions, as explained in previous sections. Each participant begins with the baseline manual mode, with drivers in control of the vehicle operations. During the manual mode, the drivers are required to operate basic driving events such as surpassing, accelerating/decelerating, etc. Then, they undertake the two modes of autopilot driving (L3 autonomous driving level), including both standard and personalized versions\footnote{Determined by participants' behaviors in manual mode.}. The two auto-driving versions are completed in a counterbalanced order. During the study, driving statistics are logged. After each driving session, the participants were asked to fill out a questionnaire about the cognitive aspects of their experience, including an assessment of perceived trust, comfort and situation awareness. Note that these are not collected for the BROOK dataset, and it's only available upon request. The study of each participant lasts approximately one hour.

\section{Example Studies of BROOK dataset}
\label{sec:example}

To justify the potential application fields and the feasibility of our design purposes, we have conducted three example studies by leveraging the BROOK dataset. In the first case study, we have built an in-vehicle multi-modal predictor to estimate drivers' physiological data via facial expressions only. Our second case study  is the first attempt to demystify the complex interactions between drivers and vehicles, by utilizing a novel self-clustering algorithm. Our third case is the first attempt to characterize and study the feasibility and quality of Differential Privacy-enabled data protection methods in IoV. Through these three example studies, we are confident that BROOK could both support real-world applications, and stimulate potential research directions in HVI (and possibly beyond).
\subsection{Case 1: Unobtrusive Predictors for Multi-modal \& Driving statistics}
\label{sec:face2multimodal}

Our first example study, supported by the BROOK dataset, is a non-trivial approach to collect in-vehicle drivers' statistics, including both multi-modal and driving statistics. Since directly equipping biosensors degrades both the safety and user experiences, we propose to leverage Convolutional Neural Networks for accurate predictions. We experimentally demonstrate its potential, as proof-of-concept. We first elaborate on the motivations of this design. Then we give an overview of the overall system. Next, we illustrate our considerations of key design choices. Finally, we present empirical results and takeaways from the prototype systems.

\subsubsection{Motivations}


Collecting statistics from drivers and passengers is an essential building block for adaptive and personalized Human-Vehicle Interactions. However, directly equipping biosensors causes potential issues, in terms of both safety and user experiences, during the driving procedure. Specifically, (1) obtaining eye-tracking data from drivers by eye detector may block driver's view, which accounts for serious safety issues \cite{DBLP:conf/etra/BlignautB12a}; (2) using bio-sensors to collect heart rate from a driver will highly affect the comfort of driving. In a real-world scenario, continuously retrieving user data via bio-sensors is impractical, user-unfriendly and cost-ineffective. Therefore, we envision the potentials of unobtrusive techniques for intelligent HVI systems, by taking only facial records to estimate both drivers' physiological statistics for monitoring drivers' status and contextual data for better run-time decision-making. We validate such approaches using state-of-the-art Convolutional Neural Networks to predict three types of driving statistics (i.e. Heart Rate (HR), Vehicle Speed (VS) and Skin Conductance (SC)), and present empirical results and takeaways.

\subsubsection{System Overview: Detect, Predict and Visualize}
The first component of our system is  \textbf{Face Detectors}. We use in-vehicle cameras to record drivers' facial videos, and leverage OpenCV library \cite{bradski2008learning} to detect driver faces and cut these videos into consecutive frames. For further processing, we resize all frames into 224 x 224 pixels as the regularized input size for further components.

The second component of our system is the \textbf{Convolutional Neural Network Models}. We apply DenseNet~\cite{DenseNet} as the core model of this component for drivers' multi-modal predictors. The input of DenseNet is the pre-processed drivers' facial images, and the output is the estimated values of our CNN. We have adjusted several parameters to improve the performance of predicting drivers' status (e.g. 100 layers instead of the suggested number of layers in the vanilla model). Table~\ref{DenseNetDetails} shows the detailed parameters of our models. The train, test and evaluation set is split in a ratio of 8:1:1, with the support of 22 drivers' facial videos labeled with HR, VS and SC from BROOK.

\begin{figure*}[!h]
\minipage{0.48\textwidth}
  \centering
  \includegraphics[width=0.9\linewidth]{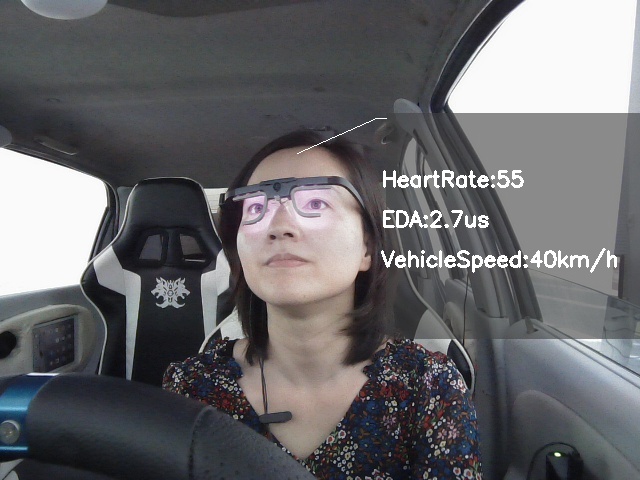}
  \caption{Demonstration of \textbf{unobtrusive techniques for collecting drivers' statistics}. Note EDA refers to Skin Conductance.}~\label{estimator}\endminipage\hfill
\minipage{0.48\textwidth}%
  \includegraphics[width=\linewidth]{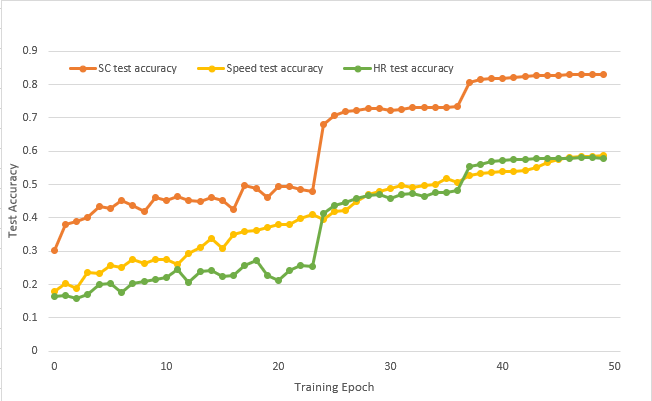}
  \caption{Test accuracy of Skin Conductance,Heart Rate and Vehicle Speed.}~\label{accuracy}
\endminipage\hfill
\end{figure*}

The third part component of our system is \textbf{Visualization Support}. Our engineering efforts enable real-time predictions of Skin Conductance, Heart Rate and Vehicle Speed. Hence, we also achieve real-time visualization of these predictions. Figure \ref{estimator} shows how our predictions are visualized across all frames.

\begin{table*}
    \caption{Major Parameters about DenseNet for BROOK in details.}~\label{DenseNetDetails}
    \resizebox{150mm}{5mm}{
    \begin{tabular}{|c|c|c|c|c|c|c|c|}
    \hline
    Parameters & Depth/Layers & Growth Rate & Dense Blocks & Compression Factor & Batch Size & Initial Learning Rate & Training Epochs \\ \hline
Values     & 100          & 12          & 4            & 0.5                & 128        & 0.1                   & 50   \\
    \hline
    \end{tabular}
    }
\end{table*}

\subsubsection{Design Choices}

We apply DenseNet\cite{DenseNet} as the architecture of the Convolutional Neural Networks (CNN) because: compared with other models, DenseNet directly applies cross-layer feature reuses, which substantially reduces the complexity of parameter tuning and adjustments. Since the aim of our study is to provide the proofs-of-concept, we believe the explorations of suitable models (e.g. MobileNet \cite{mobilenet}) are essential to identify the characteristics of such scenarios and hopefully stimulate novel designs.

\subsubsection{Empirical Results}

We report our prototype system results as follows. Our prototype systems are trained and validated, with the support of the BROOK dataset. The validation accuracy of Skin Conductance, Heart Rates, and Vehicle Speed are 82.96\%, 57.90\%, and 58.75\%. We provide more concrete details in Figure \ref{accuracy}, by setting different training epochs. In the midterm of training (epoch 25), we reduced the learning rate gradually, and we could notice that for SC and HR, they converged steeply after learning rates change. Overall, our key observation hereby is that the estimation of Skin Conductance (SC) is much higher than others. We believe this is because the granularity of SC recordings, in terms of timeline, is much smaller than others.

\subsection{Case 2: Applying self-clustering Methods for Adaptive Driving Style Characterizations}
\label{sec:dbscan}

The second case study focuses on quantitatively classifying complicated interactions between driving styles and driving behaviors adaptively. Conventional perspectives, through static partitions of driving styles, might simplify the classifications of driving styles. Our goal is to obtain diverse driving styles, derived from adaptive classifications of driving behaviors. To this end, we apply self-clustering algorithms to examine adaptive classification of driving styles, and how driving behaviors and styles interact in this case. We first elaborate on our motivation for this study. Then we provide the detailed methodology of this study. Next, we introduce the key design choices of this study. Finally, we present key observations from the experimental results and discuss major takeaways.

\subsubsection{Motivations}

Driving styles are sets of classified driving behaviors, based on relevant activities and their measurements. According to an early summary \cite{MDSI}, driving styles are classified into eight main categories: dissociative, anxious, risky, angry, high-velocity, distress reduction, patient, and careful. Though such classifications of different driving styles are already complicated (i.e. including the Choices of Speed, Brake Status, and etc\cite{DrivingBehavior}), variations between different time intervals have been generally overlooked. Moreover, the methodologies and mechanisms to enable adaptive classifications of driving styles remain understudied. Our goal is to study such variations via a novel perspective, by equalizing clusters as driving styles. To relax the constraints of conventional approaches (i.e. predetermined number of styles), we introduce Self-Clustering Algorithms as the classifications of Driving Styles. We proof-of-concept such approaches using DBSCAN Self Clustering algorithm, and present empirical observations and takeaways.

\subsubsection{Our Methodology: Adaptive Classifications via Self-Clustering Algorithms}
The driving style characterization is mainly determined by vehicle statistics in different time spots. In our experiment, we have chosen most of the vehicle data in the BROOK dataset as the source input, such as Vehicle Acceleration Rate and Steering Wheel positions. We first \textbf{\underline{pre-process Data}} to reduce the noisy points for further clustering. In this stage, we apply three main methods and we elaborate them as follow: (1) we eliminate the start-up and parking stages from all data records since our focus is to analyze the driving styles and behaviors under a stable and consecutive driving procedure; (2) we apply Min-Max Normalization\cite{MinMax} to regularize all statistics for a uniform format across all data streams, since the primary units and scales of the different kinds of data streams may vary greatly; and (3) we apply Principal Component Analysis (PCA) \cite{PCAtutotial} to retrieve principal components from the regularized, multi-dimensional data streams.

We then perform \textbf{\underline{Self-Clustering via DBSCAN Algorithm.}} To traverse all potential driving styles, we apply DBSCAN, a representative self-clustering algorithm, to enable adaptive classifications of driving styles, according to relevant driving behaviors. DBSCAN finds core samples of high density in the given dataset and expands clusters from them. Unlike conventional approaches like K-Means, DBSCAN doesn't require a predetermined number of clusters but only takes Distance functions and Parameter Settings as input parameters.

\subsubsection{Design Choices}

Since DBSCAN can accept multiple kinds of Distance Functions, hereby we first elaborate our key design choices in terms of it. There are many kinds of widely-used distance functions, such as Euclidean Distance, Manhattan Distance, etc. Since our study aims to validate the potentials of such approaches, we use Euclidean Distance as our attempt. We believe it's promising to vary different Distance Functions and validate their differences, which can substantially deepen the computational understanding of relationships between driving styles and behaviors.


As for Parameter Settings, DBSCAN takes \textit{minPoints}, \textit{epsilon} as parts of Parameter Settings: (1) \textit{minPoints} refers to the minimum numbers of data points within a cluster; and (2) \textit{epsilon} (eps) refers to the max radius of the cluster. In this case, we follow the recommended settings and vary them slightly to observe their differences. As Table~\ref{parameters} shows, we attempt nine different combinations of relevant settings. We scale \textit{eps} from 0.125 to 0.5, and scale \textit{minPoints} from 1 to 9. Since the results suggest there are negligible differences within our clustering results, we leverage representative cases, in some scenarios, and perform the analysis.

\begin{table*}
\caption{Groups of parameters and the corresponding clustering results.} ~\label{parameters}
\begin{tabular}{|c|c|c|c|c|c|c|c|c|c|} \hline
eps              & 0.125 & 0.125 & 0.125 & 0.25 & 0.25 & 0.25 & 0.5 & 0.5 & 0.5 \\  \hline
minPts           & 1     & 6    & 9   & 1   & 6     & 9   & 1   & 6   & 9   \\  \hline
clusters numbers & 8   & 4    & 3   & 6 & 4   & 3  & 6   & 4   & 3   \\  \hline
\end{tabular}
\end{table*}

\subsubsection{Empirical Results \& Observations}
\begin{figure}
    \subfigure[]{
        \centering
        \includegraphics[width=0.3\linewidth]{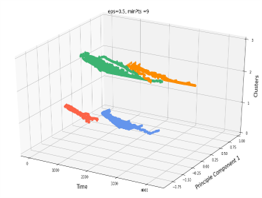}
        \label{3dResult}
    }
    \subfigure[]{
        \centering
        \includegraphics[width=0.3\linewidth]{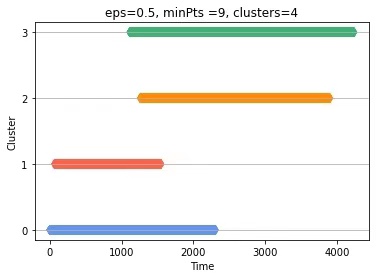}
        \label{time-cluster mapping}
    }
    \subfigure[]{
        \centering
        \includegraphics[width=0.3\linewidth]{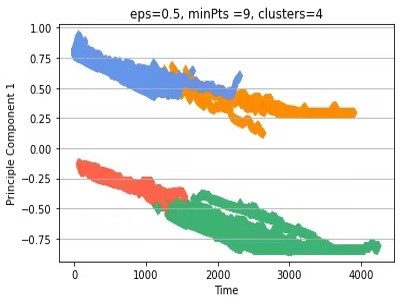}
        \label{time-pca mapping}
    }
    \centering
    \caption{Sub-Figure~\ref{3dResult} present the 3D clustering results of all drivers (i.e. eps=0.5 and minPts=9); Sub-Figure~\ref{time-cluster mapping} is the mapping in Time-Cluster plane from Sub-Figure~\ref{3dResult}; Sub-Figure~\ref{time-pca mapping} is the mapping in Time-Feature plane from Figure \ref{3dResult}.} 
    
\end{figure}

We introduce details about Sub-Figure~\ref{3dResult}:  the x-axis refers to Time, the y-axis refers to Feature (i.e. the most Principal Component) and the z-axis refers to the clusters. Sub-Figure~\ref{time-cluster mapping} and ~\ref{time-pca mapping} presents relevant 2D planes from the presented 3D figure. We make the following two observations.

\noindent
\textbf{Observation 1: Feasibility for Temporal Characterizations and Adaptive Mechanisms for Driving Styles.} Figure~\ref{3dResult} shows the three-dimensional clustering results of all drivers when DBSCAN parameters \textit{eps}, \textit{minPts} are 0.5 and 9 respectively. We obtain two facts. First, the combinations of all clusters cover the whole timeline. Second, Sub-Figure \ref{time-pca mapping} confirms such coverage can be achieved through only the most significant feature, without including every stream of multi-modal statistics. These facts show us the need of Temporal Characterizations and Adaptive Mechanisms for Driving Styles.

\noindent
\textbf{Observation 2: Classifications of Driving Styles can be Non-Deterministic.}
Sub-Figure~\ref{time-cluster mapping} shows the results when the plane of Time-Cluster is presented in 2D manner. We obtain the fact that there are overlaps between different driving styles for the specific driver. This fact validates our hypothesis that the conventional partition of driving styles may not reflect these characteristics well.

\begin{figure}[h]
    \subfigure[]{
        \centering
        \includegraphics[width=0.4\linewidth]{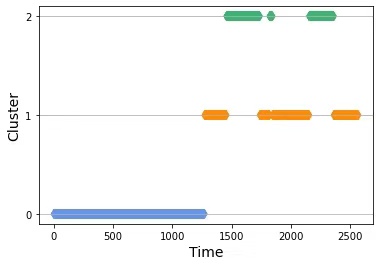}
        \label{2D mapping-tc-1}
    }
    \subfigure[]{
        \centering
        \includegraphics[width=0.4\linewidth]{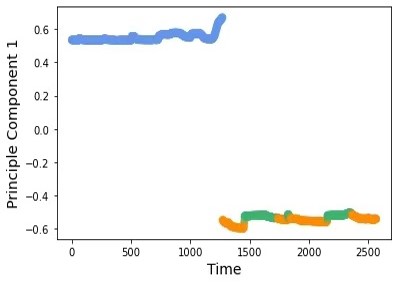}
        \label{2D mapping-tp-1}
    }
    
    \subfigure[]{
        \centering
        \includegraphics[width=0.4\linewidth]{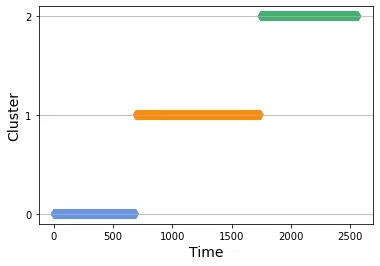}
        \label{2D mapping-tc-2}
    }
     \subfigure[]{
        \centering
        \includegraphics[width=0.4\linewidth]{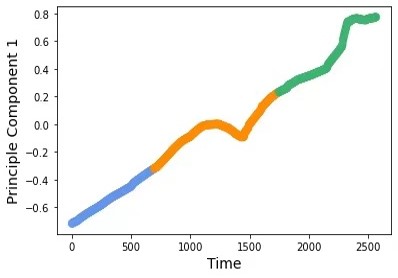}
        \label{2D mapping-tp-2}
    }
    \centering
    \caption{Concrete Examples to Demystify the Interactions between Driving Styles and Behaviors. Sub-Figures~\ref{2D mapping-tc-1} and ~\ref{2D mapping-tp-1} present one example driver set; and  Sub-Figures~\ref{2D mapping-tc-2} and ~\ref{2D mapping-tp-2} present the other example driver set.}
    \label{2D mapping}
\end{figure}

\noindent
\textbf{Observation 3: Personalized Classifications of Driving Styles are Possible.}
We use concrete examples to elaborate the possibilities of personalized classifications of driving styles. Figure~\ref{2D mapping} shows the plane of Time-Cluster from 3D figures, in a breakdown manner. We compare two example driver sets in detail. Sub-Figures~\ref{2D mapping-tc-1} and ~\ref{2D mapping-tp-1} represent one driver set, and Sub-Figures~\ref{2D mapping-tc-2} and ~\ref{2D mapping-tp-2} represent the other one. We obtain the fact that these two driver sets have different driving styles, and the differences are significant. The first driver set (Sub-Figures~\ref{2D mapping-tc-1} and ~\ref{2D mapping-tp-1}) has more continuous and smooth driving-style transitions. But the second driver set (Sub-Figures~\ref{2D mapping-tc-2} and ~\ref{2D mapping-tp-2})  has more fluctuate and frequently-switched driving-style transitions.

\subsection{Case 3: Differential Privacy for the Internet-of-Vehicles}

Our third study aims to study the tradeoffs between different Differential Privacy (DP) approaches in the context of Internet-of-Vehicles (IoV). Data privacy is a major concern with the emergence of IoV. However, it's still unclear how to deploy efficient privacy-preserving methods for emerging IoV applications. Our goal is to characterize modern privacy-preserving techniques, namely Differential Privacy (DP), for data processing in IoV. We first elaborate on our motivations for this study. Then we provide our methodology of these studies for data protection using DP. Next, we introduce the design choices of this study. Finally, we summarize the key findings from the results.

\subsubsection{Motivations}
IoV is a large-scale distributed network that consists of individual vehicles as nodes. There are diverse and frequent communications within IoV, by orchestrating humans, vehicles, things and environments, with vehicles as the interface. Such a nature of frequent communications and synchronizations, which is the same as the distributed systems, incurs the concerns of privacy issues for data processing (e.g. centralized servers). To resolve such issues, Differential Privacy has been widely-used, which is a mathematically-guaranteed mechanism for protecting private data. However, it's still unclear regarding the tradeoffs when applying such techniques in the context of IoV. To this end, we leverage the BROOK dataset to characterize the impacts of DP mechanisms within the era of IoV.

\subsubsection{Techniques: Differentially-Private Data Protection}
The \\$\epsilon$-Differential Privacy is achieved by adding random Laplace noises into the dataset. A mathematical expression of a randomized function $N_f$ could be denoted as follows.

\begin{theorem}
    (Randomized Function subject to $\epsilon$-Differential Privacy) We regard a randomized function $K_f$ as satisfying $\epsilon$-Differential Privacy, assume that $\forall$ dataset $D_1$ and $D_2$ with at most one element difference, and $\forall S \subseteq Range(K_f)$, the $K_f$ function obey the below in-equation:\\
    \centerline{$Pr[K_f(D_1) \in S] \leq exp(\epsilon) \times Pr[K_f(D_2) \in S]$}
\end{theorem}
Note that the value of $\epsilon$ is in inverse correlation with the extent of randomization (i.e. small $\epsilon$ values indicate powerful randomization).


\subsubsection{Study Design}
We focus on how to leverage DP to enable privacy-preserving data processing. To achieve this goal, we first build our DP-protected database, with the support of BROOK. We leverage the Heart Rate and Skin Conductivity data in BROOK to construct this toy database, and add random noises to the query operators via DP. In this manner, querying driving statistics in the toy database could be viewed as data processing in IoV. Second, we apply four state-of-the-art query mechanisms (i.e. Fourier \cite{07_SIGMOD_fourier}, Wavelet \cite{10_IEEE_Wavelet}, Datacube \cite{11_SIGMOD_datacube}, and Hierarchical \cite{hierarchical}) to perform a list of queries from real-world workloads. To simulate as many representative queries as possible, we consider both All-Range Query and One-Way Marginal Query. Third, we use Absolute Error and Relative Error compared with the queries on raw data as the evaluation metrics to assess DP-enabled processing on sensitive data.

\noindent
\textbf{\\DP Mechanisms for processing:}
First, we will introduce the four state-of-the-art query mechanisms: 1) Fourier is designed for workloads having all K-way marginals, for any given K. It transforms the cell counts, via the Fourier transformation, and obtains the values of the marginals based on the Fourier parameters; 2) Wavelet is designed to support multi-dimensional range workloads. It applies the Haar wavelet transformation to each dimension; 3) Datacube chooses a subset of input marginals to minimize the max errors of the query answer, to support marginal workloads; 4) Hierarchical is built for answering range queries. It uses a binary tree structure of queries for such a  goal. The first query is the sum of all cells and the rest of the queries recursively divide the first query into subsets.

\noindent
\textbf{\\Dataset Setup and Query Types:}
To characterize the query errors of DP-enabled processing, we handcraft our dataset with the support of BROOK and import it into our simulation platform\footnote{We emulate IoV via the remote APIs of modern distributed database systems. This ensures our characterizations not to mistake any procedure, by overlooking the impacts of IoV.}. In this experiment, we use the Heart Rate and Skin Conductivity as a proof-of-concept. Assume the data table is of size $m \times n$, and the constraint condition for cell $(i, j)$ could be written as $\rho_{ij}$. Then, a $length-(m \times n)$ column vector $\mathbf{x} = (x_{00}, x_{01}, ... , x_{(m - 1)(n - 1)})^T$ where $x_{ij}$ meets the condition $\rho_{ij}$ could be extracted from the table. In this case, it is equal to $(12, 17, 0, 0, 2254, ..., 11)^T$, a column vector with length of 32. With the existence of this data vector $\mathbf{x}$, it is convenient to answer a linear query. 

To simulate real-world query types, we leverage \textbf{All-Range Query} and \textbf{One-Way Marginal Query} in the query workloads. The dot product of a row vector with all positions filled with one and an equal-length column data vector, is the expected result of All-Range Query. The expected results of One-Way Marginal Query on an n-dimension database is a vector whose each position corresponds to the counting result of a query related to one attribute only. The length of the returned vector is equal to the magnitude of all possible settings of the selected attribute. Note that One-Way Marginal Query is a special case of K-Way marginal Query when $K=1$.

\noindent
\textbf{\\Evaluation Metrics and Configurations of Workloads:}
We introduce two evaluation metrics to assess query errors respectively, which are Absolute Error and Relative Error. Absolute Error refers to the absolute differences between returned value by DP-enabled query approaches and real value queried on raw data. Note that when query results of DP-enabled queries $r$ and raw query $r'$ are all vectors, the absolute error is $\|r'-r\|_1$. Else, the absolute error is $|r'-r|$. Relative Error is the ratio of absolute error to the real value. The relative error is $|(r'-r)/r|$ when expressed in the above example.

The handcrafted dataset is in size of $4\times 8$, with different $\epsilon$ values (e.g.  0.1, 0.2, 0.5, 1 and 2.5) for data protection. As for One-Way Marginal Query, we rigorously testify the query results in various dimensions (i.e. $32$, $4\times8$, $4\times4\times2$, $4\times2\times2\times2$ and $2^5$).

\subsubsection{Empirical Results \& Observations}

\begin{figure}[htp]
\centering

    \begin{minipage}{0.98\columnwidth}
        \centering
        \includegraphics[width=0.98\columnwidth]{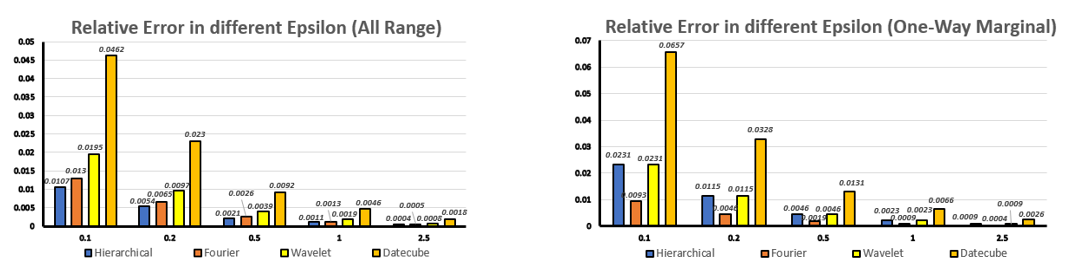}
        \caption{Relative Error of Four Query Methods}
        \label{fig:relative}
    \end{minipage}

    \begin{minipage}{0.98\columnwidth}
        \centering
        \includegraphics[width=0.98\columnwidth]{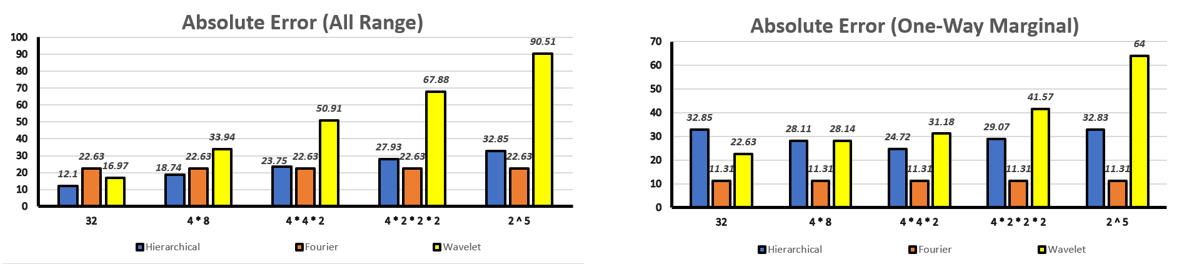}
        \caption{Absolute Error of Four Query Methods}
        \label{fig:absolute}
    \end{minipage}

\label{fig:}
\end{figure}
Figure~\ref{fig:relative} and Figure~\ref{fig:absolute} report the Relative Error and Absolute Error of the above four mechanisms for DP-enabled data sharing. We make the following two key observations.
\noindent
\textbf{\\Key Observation 1: different DP-enabled mechanisms achieve different levels of Relative Error.} Figure~\ref{fig:relative} reports the query results of Relative Errors for both All-Range Query and One-Way Marginal Query. We draw two observations. First, Datacube produces more errors than the rest mechanisms in all scenarios, for both All-Range Query and One-Way Marginal Query, which suggests it may not be an ideal choice in practice. Second, Hierarchical is the optimal method in All-Range Query, but Fourier outperforms it in One-Way Marginal Query, for any $\epsilon$ settings.

\noindent
\textbf{\\Key Observation 2: different DP-enabled mechanisms achieve different levels of Absolute Error.} Figure~\ref{fig:absolute} shows the query results of Absolute Errors for both All-Range Query and One-Way Marginal Query. We draw four observations. First, Hierarchical and Wavelet both incur more errors when the granularity of query structures grows in both query types. Second, Fourier is the most robust mechanism when the granularity of query structures grows in both query types. It also achieves the best results in complex granularity of in All-Range Query and also the best in all the settings of One-Way Marginal Query. Third, Hierarchical obtains better results from query structure $32$ and $4 \times 8$ in All-Range Query. Four, Wavelet can achieve fewer query errors compared with Hierarchical with a small granularity of queries in One-Way Marginal Query.

\section{Discussions}
\label{sec:futureresearch}

BROOK dataset is only a starting step to provide comprehensive data support for the research of Human-Vehicle Interaction systems. Though there are certainly several limitations of the BROOK dataset, we believe the BROOK dataset also delivers a lot of takeaways in terms of future research. In this section, we discuss both limitations and takeaways from the BROOK dataset and its example studies. We first outline several key limitations of the BROOK dataset (Section~\ref{subsec:limitations}), and then elaborate key takeaways from the BROOK dataset and its example studies (Section~\ref{subsec:takeaways}).

\subsection{Limitations of BROOK dataset}
\label{subsec:limitations}

Though there are some novel features of the BROOK dataset, it still has several limitations. We summarize them into three aspects as follows. First, the BROOK dataset is collected under in-lab simulations, instead of in-field studies. This is because building a comprehensive dataset is impractical from real-world studies, where certain restrictions prevent us from equipping as many devices as possible for drivers; second, the BROOK dataset only consists of fully-automated mode and manual mode, and the collections for semi-automated mode are still required to provide thorough coverage. This is because the rationale to provide a semi-automated mode for individual drivers is still unclear; Third, the BROOK dataset is restricted to drivers from a specific region (i.e. China), because a broader coverage of participants requires a significant amount of effort and resources.

We believe the efforts to build such a dataset are far beyond the capability of a single research group. Therefore, we position the BROOK dataset as a starting point, and we hope the BROOK dataset can bring important learning lessons for future efforts. Moreover, we envision the BROOK dataset can stimulate more follow-up works and collaborations, to achieve a broader coverage in terms of driving modes, regions, and in-field studies.

\subsection{Takeaways from BROOK dataset \& Example Studies}
\label{subsec:takeaways}
The empirical results from our example studies suggest the high quality of the BROOK dataset. Therefore, we believe BROOK is capable to support multiple research purposes, and provide data sources for evaluations/validations. Particularly, we believe BROOK is sufficient for examining the suitability of diverse data-driven techniques and interactive designs in the context of Human-Vehicle Interaction systems. We discuss takeaways of our example studies in detail.

In our first example study, we proof-of-concept the effectiveness of our designs. We believe future works can enable more accurate in-vehicle multi-modal Predictors. To build a personalized Human-Vehicle Interaction (HVI) system, multi-modal streams are critical sources in the decision-making module of HVI systems. Particularly, using multiple sensors to collect drivers' statistics in vehicles affects safety and user experiences, which substantially limits its practicality. Though our works already demonstrate the potential, we believe there are still rooms for more powerful predictors(e.g. more detailed facial expression \cite{FE1,FE2, automotive21/facial-expressions}) and customization-friendly predictors~\cite{hci22/face2statistics} inside vehicles. It also of great pracical values to integrate them into certain platforms (e.g., IoV Emulation Platform \cite{automotiveui21/iov}) to improve the overall Quality-of-Service.

In our second example study, we identify the complicated interactions between driving behaviors and styles, and we believe there are needs for both long- and short-term characterizations of driving styles. Current studies have paid a lot of attention to the characterizations and classification of driving styles, but few attempts to adapt the instant reflections from drivers/passengers. We believe BROOK offers a great many chances for taking an in-depth analysis and modeling of drivers/passengers in both Long and Short Term (e.g. Long Short-Term Memory (LSTM) \cite{LSTM}). Despite driving styles from long-term inference, future HVI systems could be capable of providing real-time reflections on drivers'/passengers' affective states, to further improve its quality of service. Thus, characterizing driving styles in a finer granularity is essential, to enable adaptive and personalized Human-Vehicle Interactions.

In our third example study, we characterize the effects of leveraging Differential Privacy mechanisms on data processing and data sharing within IoV. Data privacy will be a major concern when IoV systems grow mature. Previous works utilize blockchain-based mechanisms (e.g. \cite{blockchain-iov}) to address the data privacy issues. However, there is still a huge gap before striking a balance between sensitive data utility and privacy in the context of IoV. Previous works~\cite{UCC-TR/DP} provides 11 key guidelines for applying DP-methods in the context of IoV and there exists 
some early attempts to mitigate the negative effects of DP methods (e.g., \cite{UCC-TR/HUT}). We believe BROOK and our study could pave the road for such a goal and stimulate more follow-up works. We have already narrowed down the $\epsilon$ values to avoid redundant future works, and conclude the strength and weakness of introducing DP-enabled protections in practice.

\section{Conclusions}

We present BROOK dataset, a multi-modal and facial video dataset to exploit data-driven techniques for adaptive and personalized Human-Vehicle Interaction designs. The current version of BROOK contains 34 drivers' driving statistics under both manual and autonomous driving modes. Through BROOK dataset, we conduct three example studies, via advanced data-driven techniques, to demonstrate its potentials in practice. Our results justify the applicability and values of the BROOK dataset. We believe, with the emerging trends of Autonomous Vehicles, more efforts on similar datasets are essential to unleash the design spaces.


\bibliographystyle{ACM-Reference-Format}
\bibliography{sample-base}

\end{document}